\documentclass[twocolumn,preprintnumbers,amsmath,amssymb]{revtex4}
\usepackage{array}
\usepackage{booktabs}
\usepackage{tabu}
\usepackage{dcolumn}
\usepackage{amsmath}
\usepackage{amsfonts}
\usepackage{amssymb}
\usepackage{graphicx}
\usepackage[colorlinks,linkcolor=red,citecolor=blue]{hyperref}
\usepackage{subfigure}
\usepackage{graphicx}
\usepackage{dcolumn}
\usepackage{bm}
\newtheorem{thm}{Theorem}
\newcommand{\bra}[1]{\langle #1 \vert}
\newcommand{\ket}[1]{\vert #1 \rangle}
\newcommand{\commentold}[1]{}
\DeclareMathSymbol{:}{\mathpunct}{operators}{"3A}

\def\be{\begin{equation}}
\def\ee{\end{equation}}
\def\bea{\begin{eqnarray}}
\def\eea{\end{eqnarray}}
\def\f{\frac}
\def\n{\nonumber}
\def\l{\label}

\begin{document}

\title{Quantum Thermodynamic Force and Flow}
\author{B. Ahmadi}
\affiliation{Department of Physics, University of Kurdistan, P.O.Box 66177-15175, Sanandaj, Iran}
\author{S. Salimi}
\email{shsalimi@uok.ac.ir}
\affiliation{Department of Physics, University of Kurdistan, P.O.Box 66177-15175, Sanandaj, Iran}
\author{F. Kheirandish}
\affiliation{Department of Physics, University of Kurdistan, P.O.Box 66177-15175, Sanandaj, Iran}
\author{ A. S. Khorashad}
\affiliation{Department of Physics, University of Kurdistan, P.O.Box 66177-15175, Sanandaj, Iran}

\date{\today}

\def\br{\biggr}
\def\bl{\biggl}
\def\Br{\Biggr}
\def\Bl{\Biggl}
\def\be\begin{equation}
\def\ee{\end{equation}}
\def\bea{\begin{eqnarray}}
\def\eea{\end{eqnarray}}
\def\f{\frac}
\def\n{\nonumber}
\def\l{\label}

\keywords{Suggested keywords}
\maketitle
\newpage
\textbf{Abstract}
\newline
Why do quantum evolutions occur and why do they stop at certain points? In classical thermodynamics affinity was introduced to predict in which direction an irreversible process proceeds. In this paper the quantum mechanical counterpart of classical affinity is found. It is shown that the quantum version of affinity can predict in which direction a process evolves. A new version of the second law of thermodynamics is derived through quantum affinity for energy-incoherent state interconversion under thermal operations. we will also see that the quantum affinity can be a good candidate to be responsible, as a force, for driving the flow and backflow of information in Markovian and non-Markovian evolutions. Finally we show that the rate of quantum coherence can be interpreted as the pure quantum mechanical contribution of the total thermodynamic force and flow. Thus It is seen that, from a thermodynamic point of view, any interaction from the outside with the system or any measurement on the system may be represented by a quantum affinity.

\textbf{Background}
\newline
In classical physics, motion is explained by the Newtonian concept of force, but what is the 'driving force' that is responsible for quantum state transformations? Why do quantum evolutions occur at all and why do they stop at certain points? In classical thermodynamics, chemists proposed the same question concerning chemical reactions. Chemists called the 'force' that caused chemical reactions \textit{affinity}. The thermodynamic formulation of affinity as we know it today is due to Th\'{e}ophile De Donder (1872-1957), the founder of the Belgian school of thermodynamics. He formulated chemical affinity on the basis of chemical potential \cite{1}.
\newline
Clausius considered irreversible processes as an integral part of formulating the second law of thermodynamics. He included irreversible processes explicitly into the formalism of entropy by dividing entropy into two parts \cite{2}: the change in entropy due to the exchange of heat with the environment by the term $dQ/T$ (which is compensated by equal gain or loss of heat by the environment) and the entropy produced by irreversible processes within the system (the uncompensated transformation) $d_iS$. On the other hand, irreversible processes can in general be thought of as 'thermodynamic forces' driving 'thermodynamic flows'. The thermodynamic flows are a consequence of the thermodynamic forces. For example, the temperature gradient is the thermodynamic force that causes an irreversible flow of heat; similarly, a concentration gradient is the thermodynamic force that causes the flow of matter. Consider the free expansion of a gas, the irreversible increase in entropy of the gas is given by \cite{2}
\begin{equation}\label{1}\nonumber
d_iS=\dfrac{P_{gas}-P_{piston}}{T}dV,
\end{equation}
where P, T and V are the pressure, temperature and volume of the gas, respectively. In this case, the term $(P_{gas}-P_{piston})/T$ corresponds to the thermodynamic force and dV/dt the corresponding flow. The term $(P_{gas}-P_{piston})dV$ may be identified as the 'uncompensated heat' of Clausius.
In his pioneering work on the thermodynamics of chemical processes, De Donder incorporated the uncompensated transformation or uncompensated heat of Clausius into the formalism of the Second Law through the concept of affinity \cite{1}. He took the uncompensated heat of Clausius in the context of chemical reactions and defined the affinity of a chemical reaction, which allows to write the entropy production of the reaction in an elegant form, as the product of a thermodynamic force and a thermodynamic flow. For a chemical reaction $X + Y\rightleftharpoons2Z$, he defined a new state variable called affinity as \cite{2}
\begin{equation}\label{2}
A\equiv\mu_X+\mu_Y-2\mu_Z,
\end{equation}
where the coefficients $\mu_k$ are called the chemical potentials. This affinity is the driving force for chemical reactions. In terms of affinity $A$, the rate of increase of entropy production, $d_iS$, is written as \cite{2}
\begin{equation}\label{eq:3}
\dfrac{d_iS}{dt}=\big(\dfrac{A}{T}\big)\dfrac{d\xi}{dt}.
\end{equation}
Thus the entropy production due to chemical reactions  is a product of a thermodynamic force $A/T$ and a thermodynamic flow $d\xi/dt$. The flow in this case is the conversion of reactants to products (or vice versa), which is caused by the force $A/T$. The thermodynamic flow $d\xi/dt$ is referred to as the velocity of reaction or rate of conversion. there is no general relationship between the affinity and the velocity of a reaction. The sign of affinity can be used to predict the direction of reaction. If $A>0$, the reaction proceeds to the right and if $A<0$, the reaction proceeds to the left \cite{2}.
\newline
In chemical physics and physical chemistry, the affinity is the \textit{tendency} of a chemical species such as an atom or molecule to react with another to form a chemical compound. Therefore the main aim of our work is to find the quantum mechanical counterpart of this (classical) affinity and ensure that it has all its classical properties. In order to this we will show that, as in classical thermodynamics, in quantum thermodynamics the entropy production of a system can be expressed as the product of a thermodynamic force and a thermodynamic flow. We then ensure that the quantum affinity, as classical affinity, acts as a force which pushes an initial state to a final state thus determining in which direction a quantum process proceeds, i.e, the quantum affinity is the tendency of a system to go from a state $\rho$ to another state $\sigma$ under a quantum mechanical evolution. This means that the quantum affinity connects the arrow of time with quantum state transformation. We will also show that the quantum mechanical affinity can be considered as a good candidate to be responsible for the flow and backflow of information in Markovian and non-Markovian evolutions. Hence the quantum affinity connects also the arrow of time with the flow and backflow of information. We will finally show that the rate of the quantum coherence is the difference between the total quantum thermodynamic force and flow and the classical thermodynamic force and flow, as expected.
\newline
\newline
\textbf{Results}
\newline
\textbf{Quantum thermodynamic force and flow}. Consider an arbitrary quantum system $S$ coupled with a heat reservoir $B$ initially in thermal state at temperature $T$. The total system $S+B$ with Hamiltonian $H=H_S+H_B+H_{SB}$ is closed and thus evolves unitarily in time \cite{3, 4}. But we are primarily interested in the occurrence and characterization of irreversible behavior in the system. We thus focus our attention on the entropy $S(t)$ of the system \cite{4}, $S(t)=-tr\{\rho_s(t)\ln\rho_s(t)\}$, where $\rho_s(t)$ is the density of the state of the system which evolves from an initial state $\rho_s(0)$ to a final state $\rho_s(t)$ by completely positive and trace preserving (CPTP) maps $\Lambda_t$ \cite{3}, i.e, $\rho_s(t)=\Lambda_t[\rho_s(0)]$. The quantum thermodynamic force and flow are obtained, respectively, (see Supplementary Note 1) as
\begin{equation}\label{10}
F_{th}=\dfrac{1}{\rho_s^\beta}(\ln\rho_s^\beta-\ln\rho_s(t)),
\end{equation}
\begin{equation}\label{11}
V_{th}=\dot{\rho_s}(t)\rho_s^\beta.
\end{equation}
Thus, as in the case of entropy production due to chemical reactions in classical thermodynamics, the entropy production due to irreversible processes in quantum thermodynamics is written as a product of a thermodynamic force $\dfrac{1}{\rho_s^\beta}(\ln\rho_s^\beta-\ln\rho_s(t))$ and a thermodynamic flow $\dot{\rho_s}(t)\rho_s^\beta$. The flow in this case is the transformation of an initial quantum state to a final state. In Ref. \cite{7} $\dot V$ was introduced to describe the speed of the system evolution, where $V = Tr(\rho\rho_s)$ and $\rho_s$ is the density matrix of the target state $|S\rangle$. Notice that if the system Hamiltonian is time-dependent $\rho_s^\beta$ is replaced by $\rho_s^\beta(t)=\exp(-\beta H_s(t))/Z_s(t)$. It must be noted that in our work $\rho_s^\beta(t)$ is not the target state. It is the instantaneous equilibrium state of the system corresponding to the bath temperature $T$. And the only thing which is important about $\rho_s^\beta(t)$ in our work is that the quantum thermodynamic force vanishes in this state and it remains zero if the evolution is Markovian. But if the evolution is non-Markovian it may not remain zero because the Gibbs state may not be an invariant state of the non-Markovian map \cite{Dariusz}. Now comparing to Eq. (\ref{eq:3}), $\rho_s^\beta$  and $\ln\rho_s^\beta-\ln\rho_s(t)$ play the roles of the temperature $T$ and the affinity $A$, respectively. Therefore $F_{th}$ can be rewritten as
\begin{equation}\label{12}
F_{th}=\dfrac{A}{\rho_s^\beta}.
\end{equation}
From now on, we shall refer to $A$ as \textit{quantum} affinity and the thermodynamic flow will be referred to as the velocity of the transformation. Here we define $\bar{A}$ as
\begin{equation}\label{13}
\bar{A}(\rho)\equiv tr\{A\}(\rho).
\end{equation}
In the following we will show that, as in classical thermodynamics, the quantum affinity $A$ acts as a force and determines the direction in which the quantum processes proceed.
\newline
\textbf{Pure bipartite states}. A quantum state $\rho$ can be transformed into another quantum state $\sigma$ by LOCC (see Supplementary Note 2) if and only if
\begin{equation}\label{Eq:1}
\bar{A}(\rho)\leq\bar{A}(\sigma).
\end{equation}
Eq. (\ref{Eq:1}) shows that $\bar{A}(\rho)$ is the tendency of the state $\rho$ to go to the state $\sigma$. As in classical thermodynamics that affinity was expressed on the concept of chemical potential, $\bar{A}(\rho)$ can be interpreted as the average local non-equilibrium potential of the state $\rho$. In other words a state $\rho$ with a smaller potential is "pulled", by LOCC, toward the state $\sigma$ with a larger potential. Thus the quantum affinity $A(\rho)$ associates each state of the system with a local potential that determines whether a state can be (deterministically) transformed into another state by LOCC. We must point out that our definition of the quantum affinity is also valid for matrices with zero eigenvalues. When the eigenvalue of a density matrix goes to zero the affinity becomes larger and approaches infinity. Thus the affinity of a density matrix with a zero eigenvalue is infinity and the above statement still holds. For example if a pure bipartite entangled quantum state $\rho$ has more zero eigenvalues than another pure bipartite entangled quantum state $\sigma$ then $\bar{A}(\rho)>\bar{A}(\sigma)$. Thus the state $\sigma$ can be transformed with certainty to the state $\rho$ by LOCC which completely agrees with Nielsen's Theorem \cite{8}. Hatano and Sasa \cite{10} introduced a classical non-equilibrium potential as $\phi(x; \alpha)=-\ln\rho_{ss}(x; \alpha)$ where $\rho_{ss}(x; \alpha)$ is the probability distribution function of the steady state corresponding to $\alpha$. Similarly, Manzano \textit{et. al} \cite{11} defined $\Phi_\rho=-\ln\rho$ as the quantum non-equilibrium potential. Thus our work justifies the definition of the speed of the quantum system evolution introduced in Ref. \cite{7} and the definition of the non-equilibrium potential of the quantum system introduced in Refs. \cite{10,11}.
\newline
There exist, however, incomparable states in the sense that neither state is convertible into the other with certainty only using LOCC. Let $\ket{\psi}$ and $\ket{\phi}$ be two states with Schmidt numbers $\alpha$ and $\beta$, respectively. The transformation $\ket{\phi} \rightarrow \ket{\psi}$ is more probable than $\ket{\psi} \rightarrow \ket{\phi}$ by LOCC (see Supplementary Note 3) if and only if the $\ell$-th component of the potential difference
\begin{equation}\label{Eq:2}
\triangle A_\ell=A_\ell(\rho_\psi)-A_\ell(\rho_\phi)>0.
\end{equation}
\textbf{Quantum affinity and the Second Law}. A state $\rho$ block diagonal in energy eigenbasis can be transformed with certainty into another block diagonal state $\sigma$ by thermal operations (see Supplementary Note 4) if and only if
\begin{equation}\label{Eq:3}
\bar{A}(\hat{\alpha})>\bar{A}(\hat{\beta}),
\end{equation}
in which $\alpha, \beta$ are the probability vectors of the states $\rho$ and $\sigma$, respectively. Eq. (\ref{Eq:3}) is another way of stating the second law of thermodynamics for states block diagonal in energy: "$\bar{A}(\hat{\alpha})$ of a state $\rho$ with probability vector $\alpha$ never increases under thermal operations". Hence, quantum affinity $A(\rho)$ connects the arrow of time with quantum state transformation. In other words, the thermodynamic arrow of time always points in the direction of decreasing quantum state affinity $\bar{A}(\hat{\alpha})$ under thermal operations.
\newline
\textbf{Quantum affinity, heat and work}. The rate of the entropy production of a quantum system interacting with a reservoir initially in equilibrium at temperature $T$ can be written as
\begin{equation}\label{Eq:5}
\dfrac{d_iS}{dt}=tr\{\dot{\rho}_sA^{tot}\}-tr\{\dot{\rho}_sA^{eq}\},
\end{equation}
where the total quantum affinity was defined as $A^{tot}\equiv-\ln\rho_s(t)$ and the equilibrium quantum affinity as $A^{eq}\equiv-\ln\rho_s^\beta$. Thus the (irreversible) quantum affinity can be expressed as the difference between the total and the equilibrium quantum affinity
\begin{equation}\label{Eq:6}
A(t)=A^{tot}-A^{eq}.
\end{equation}
Using Eq. (\ref{Eq:6}) the heat could be expressed as
\begin{equation}\label{Eq:7}
d\langle Q\rangle=tr\{\rho(t+dt)\mathbb{Q}\}-tr\{\rho(t)\mathbb{Q}\},
\end{equation}
where $\mathbb{Q}=-TA^{eq}$. Hence the equilibrium quantum affinity is the force which pushes (pulls) information out of (into) the system to (from) its environment in the form of heat. Now the (irreversible) quantum affinity can be interpreted as the force which is responsible for the information exchanged, between the system and the environment, not in the form of heat. This type of information exchange occurs in the interior of the system and thus may be reused by the system to do work. For instance information may be stored in the \textit{correlations} established, during the strong interaction of the system with its environment. In the next section we will show that whenever $\bar{A}(\rho)$ begins to increase information backflows into the system. Consider the free expansion of an isolated (classical) gas of non-interacting particles. It seems like there exists a force which pushes the gas to expand (or to become more disordered). The relation $d_iS=F_{th}d\xi$, in classical thermodynamics, is similar to the relation, in classical mechanics, $dW=Fdx$. $F_{th}$ and $d\xi$ play the roles of the force $F$ and the displacement $dx$, respectively. $d_iS$ is in fact equal to $\beta dW_{irr}$ where $dW_{irr}$ is the irreversible work and is always positive in (deterministic) classical thermodynamics due to the Clausius' statement of the Second Law, thus information is always encoded which, in turn, leads to the fact that the Carnot engine is the most efficient engine. In a further publication \cite{Ahmadi} we will examine more properties and uses of the quantum affinity in quantum thermodynamics as a force. We will show that the relation $d_iS=\beta dW_{irr}$ also holds in quantum thermodynamics and the quantum affinity is responsible for encoding and decoding information. Whenever it decodes information more work, than what is expected, can be extracted from the system leading to an engine more efficient than that of Carnot. It will also be shown that Maxwell's demon \cite{Leff} in quantum thermodynamics is in fact a quantum affinity which forces the information back into the system, i.e., it decodes information. We will also reestablish the Landaure's principle \cite{Landauer} in the language of the quantum affinity. It should be pointed out that, from a thermodynamic point of view, LOCC are in fact Maxwell's demons intervening in the process and inequalities (\ref{Eq:1}) and (\ref{Eq:2}) mean that it seems like there exists a force implementing the change LOCC make on the state of the system. The details mentioned above indicate the fact that, from a thermodynamic point of view, any interaction from the outside with the system or any measurement on the system may be represented by a quantum affinity.
\newline
\textbf{Quantum affinity and non-Markovianity}. During a Markovian evolution information flows out of the system into its environment but in a non-Markovian evolution, due to correlations between the system and its environment, information backflows into the system from its environment. Here we will show that the quantum affinity is the tendency of the system to establish correlations with its environment and is the driving force responsible for the flow and backflow of information. Hence the quantum affinity connects the arrow of time with the flow and backflow of information. The quantum affinity $A(\rho_s(t))$ is a function of the map $\Lambda_t$, thus it behaves differently during Markovian and non-Markovian evolutions. The collapses and revivals of $A(\rho_s(t))$ in the first example, below, show the fact that the dynamics of the system undergoes Markovian and non-Markovian evolutions during the process. But it should be noted that $A(\rho_s(t))$ is the total thermodynamic affinity at time t. Consider a master equation with two decay rates, $\gamma_1(t)>0$ and $\gamma_2(t)<0$, that is non-Markovian at all times. If the Markovianity dominates the non-Markovianity, i.e, $|\gamma_1(t)|>|\gamma_2(t)|$ then revivals are not observed in the behavior of $\bar{A}(\rho_s(t))$, although the dynamics is non-Markovian at all times (see the second example). In order to separate the contributions of Markovianity and non-Markovianity in $\bar{A}(\rho_s(t))$ we take the time derivative of $\bar{A}(\rho_s(t))$ (see Supplementary Note 5),
\begin{equation}\label{27}
\dfrac{d\bar{A}(\rho_s(t))}{dt}=\sum_{k=1}^{d^2-1}\dfrac{d\bar{A}^k(\rho_s(t))}{dt},
\end{equation}
where
\begin{equation}\label{Eq:4}
\dfrac{d\bar{A}^k}{dt}\equiv-tr\{\gamma_k(t)[L_k(t)\rho_sL_k^\dag(t)-\dfrac{1}{2}\{L_k^\dag(t)L_k(t),\rho_s\}]\rho^{-1}_s\},
\end{equation}
Now the effect of Markovianity and non-Markovianity can be clearly seen separately. In the following examples we will illustrate how $\bar{A}(\rho_s(t))$ behaves differently during Markovian and non-Markovian dynamics and we see that if $\gamma_{k=i}(t)>0$ then $A^i$ is the force driving the flow of information and if $\gamma_{k=j}(t)<0$ then $A^j$ is the force driving the backflow of information.

\begin{figure}[h]
\centering
\includegraphics[width=8.5cm]{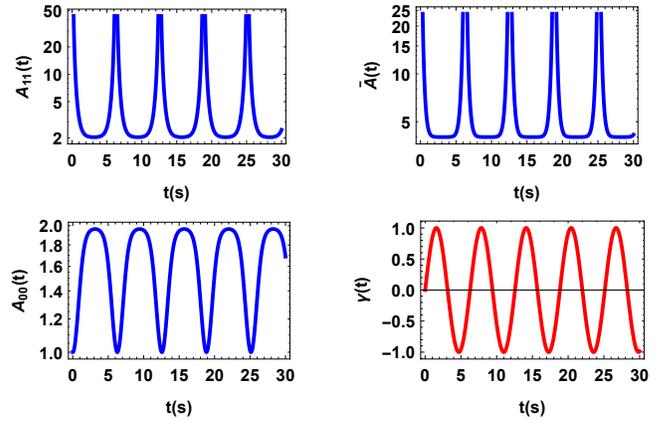}
\caption{(Color online) $A(\rho_s(t))$ vs. time t for a qubit with decay rate $\gamma(t)=\sin(t)$. As is anticipated revivals and collapses are observed that give rise to the temporary flow and backflow of information which suggest that $A(\rho_s(t))$ is the force responsible for driving the flow and backflow of information.}
\label{fig:Fig1.eps}
\end{figure}

\begin{figure}[h]
\centering
\includegraphics[width=8.5cm]{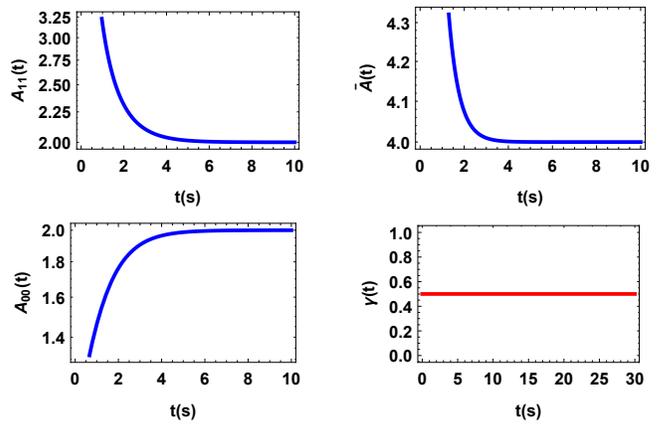}
\caption{(Color online) $A(\rho_s(t))$ vs. time t for purely dephasing dynamics with decay rate $\gamma(t)=1/2$. Since the dynamics is Markovian no revival or collapse is observed.}
\label{fig:Fig2.eps}
\end{figure}

\textbf{Examples}. Since $\ln\rho^\beta_s$ does not change over the time we neglect this term when we calculate the quantum affinity $A(\rho_s(t))$ for these examples. The following dynamical map of a two-dimensional quantum system (qubit),
\begin{equation}\label{28}
\dot\rho_s(t)=\gamma(t)[\sigma_z\rho_s(t)\sigma_z-\rho_s(t)],
\end{equation}
where $\sigma_z$ is the Pauli matrix and
\begin{equation}\label{29}
\gamma(t)=\sin(t),
\end{equation}
with $\int_{t_0}^{t_1}\gamma(s)ds\geq0$ for completely positive dynamics, is of particular interest, as it provides a simple example of a completely positive evolution \cite{16} that is Markovian when $\gamma(t)$ is positive and non-Markovian when $\gamma(t)$ is negative. As can be seen from the results plotted in Fig. (\ref{fig:Fig1.eps}) for the initial state $\rho_s(0)=\dfrac{1}{2}\begin{pmatrix}
  1 & 1 \\  1 & 1
\end{pmatrix}$, when the evolution changes its behavior from Markovian to non-Markovian the rate of $A_{ii}(\rho_s(t))$ begins to switch signs. It is also observed that for Markovian and non-Markovian evolutions $\bar{A}(\rho_s(t))$ decreases and increases, respectively, giving rise to the temporary flow and backflow of information.
As a second example consider the evolution of a qubit given by the following master equation (pure dephasing) \cite{3},
\begin{equation}\label{30}
\dot\rho_s(t)=\gamma(t)[\sigma_z\rho_s(t)\sigma_z-\rho_s(t)],
\end{equation}
where $\sigma_z$ is the Pauli matrix and $\gamma(t)=1/2$, that is Markovian during the whole evolution. For the initial state $\rho_s(0)=\dfrac{1}{2}\begin{pmatrix}
  1 & 1 \\  1 & 1
\end{pmatrix}$, a straightforward computation of quantum affinity gives that, as expected, there exist no revival and collapse in the behavior of $A_{ii}(\rho_s(t))$ and consequently no revival and collapse in $\bar{A}(\rho_s(t))$ (see illustration in Fig. \ref{fig:Fig2.eps}). Since the dynamics is Markovian at all times $\bar{A}(\rho_s(t))$ decreases (collapses) with time and never increases (revives).

\begin{figure}[h]
\centering
\includegraphics[width=8.5cm]{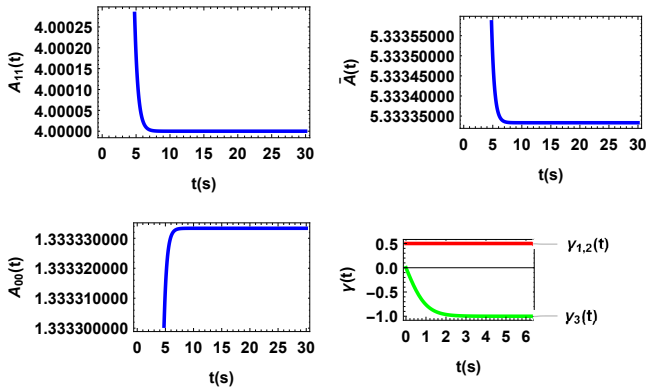}
\caption{(Color online) $A(\rho_s(t))$ vs. time t for multiply decohering dynamics with decay rates $\gamma_1(t)=\gamma_2(t)=1$ and $\gamma_3(t)=-\tanh t$. Although the dynamics is non-Markovian at all times but since the first two Markovian forces dominate the non-Markovian one no revival and collapse may appear.}
\label{fig:Fig3.eps}
\end{figure}

\begin{figure}[h]
\centering
\includegraphics[width=8.5cm]{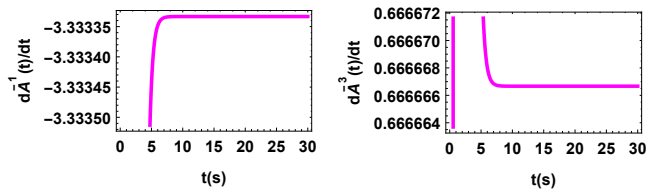}
\caption{(Color online) $\dfrac{d\bar{A}^1}{dt}$ is negative, but $\dfrac{d\bar{A}^3}{dt}$ is positive at all times showing the fact that revival (backflow) of information occurs throughout the evolution, although no revival or collapse is observed in the behavior of $A(t)$ (see Fig. \ref{fig:Fig3.eps}).}
\label{fig:Fig4.eps}
\end{figure}

We finally consider an example of multiply decohering dynamics \cite{17, 18},
\begin{equation}\label{31}
\dot\rho_s(t)=\dfrac{1}{2}\sum_{k=1}^3\gamma_k(t)[\sigma_k\rho_s(t)\sigma_k-\rho_s(t)],
\end{equation}
where the $\sigma_k$ are the Pauli $\sigma$ matrices and
\begin{equation}\label{32}
\gamma_1(t)=\gamma_2(t)=1,\ \gamma_3(t)=-\tanh t,
\end{equation}
which is non-Markovian at all times. As mentioned before, the fact that $A_{ii}(\rho_s(t))$ undergoes no collapse and revival with time (as shown in Fig. \ref{fig:Fig3.eps}) is because the first two Markovian forces dominate the non-Markovian one. As depicted in Fig. (\ref{fig:Fig4.eps}), for the initial state $\rho_s(0)=\dfrac{1}{2}\begin{pmatrix}
  1 & 1 \\  1 & 1
\end{pmatrix}$, $\dfrac{d\bar{A}^1}{dt}$ and $\dfrac{d\bar{A}^2}{dt}$ are negative but $\dfrac{d\bar{A}^3}{dt}$ is positive, during the entire evolution, which indicates that $A^3$ is responsible for the backflow of information. Considering these results it stands to reason to interpret quantum affinity $A(\rho_s(t))$ as a thermodynamic force driving the flow and backflow of information or the tendency of the system to establish correlations with its environment.

\textbf{Quantum coherence and quantum thermodynamic force and flow}. Quantum coherence is a landmark feature of quantum mechanics and has no classical counterpart. We separate the classical contribution of the total thermodynamic force and flow and show that the difference between the total and classical thermodynamic force and flow equals the rate of quantum coherence. By classical, here, we mean those elements of the density matrix which generate no quantum coherence in the state of the system. In order to derive a relation between quantum coherence and quantum thermodynamic force and flow we will employ the so-called relative entropy of coherence \cite{19}
\begin{equation}\label{eq:33}
C(\rho)=S(\rho_d)-S(\rho),
\end{equation}
where $\rho_d$ is the state obtained from $\rho$ by deleting all the off-diagonal elements. Now let $\{\ket{n}\}$ denote the eigenstates of the Hamiltonian $H$, and $p_n=\bra{n}\rho\ket{n}$ the corresponding populations. After some straightforward calculations the entropy production can be written as
\begin{equation}\label{eq:34}
\dfrac{d_iS}{dt}=\sum_{n}\dot{p}_n\ln\dfrac{p_n^\beta}{p_n}-\dot{C}(\rho_s).
\end{equation}
Thus the desired relation is
\begin{equation}\label{eq:35}
-\dot{C}(\rho_s)=tr\{\dot\rho_s\ln\dfrac{\rho_s^\beta}{\rho_s}\}-\sum_{n}\dot{p}_n\ln\dfrac{p_n^\beta}{p_n},
\end{equation}
where the first term on the right hand side is the total thermodynamic force and flow and the second term is the classical part of the total thermodynamic force and flow. Eq. (\ref{eq:35}) means that $-\dot{C}(\rho_s)$ is obtained by subtracting the classical part from the total thermodynamic force and flow. Therefore what remains is \textit{purely} quantum mechanical. This result is remarkable, because we have shown that, in the language of thermodynamic force and flow, $\dot{C}(\rho_s)$ can be interpreted as the pure quantum mechanical contribution of the total thermodynamic force and flow. Roughly speaking, $\dot{C}(\rho_s)$ is the off-diagonal contribution of the total thermodynamic force and flow. Therefore we have shown that in cases like the flow and backflow of information and coherence which are specific features of (stochastic) quantum mechanics the quantum affinity still acts as a force or tendency.

\textbf{Discussion}. We have shown that, as in classical thermodynamics, the entropy production can be written as the product of a thermodynamic force and a thermodynamic flow. The latter determines the velocity of the evolution. Comparing quantum thermodynamic force with its classical version we have derived the quantum mechanical version of affinity and proved that, as in classical thermodynamics, quantum affinity can predict in which direction an irreversible transformation occurs. This quantum affinity enabled us to associate a state with a local non-equilibrium potential such that pure bipartite entangled quantum states with smaller potentials are pulled toward pure bipartite entangled states with larger potentials, deterministically or nondeterministically, under LOCC. We have examined the behavior of quantum affinity under thermal operations and discovered a new version of the Second Law through quantum affinity such that the thermodynamic arrow of time always points in the direction of decreasing quantum affinity. we have also observed that quantum affinity can be interpreted as the thermodynamic force driving the flow and backflow of information in Markovian and non-Markovian evolutions, respectively, and illustrated this with three physical examples.  And lastly, using the concept of relative entropy of coherence, we have shown that in the language of thermodynamic force and flow the rate of quantum coherence can be interpreted as the pure quantum mechanical contribution of the total thermodynamic force and flow. Thus we have shown that, from a thermodynamic point of view, any interaction from the outside with the system or any measurement on the system may be represented by a quantum affinity.

\section{Supplementary Note 1}
The total change in the entropy $\Delta S$ of the system is divided into two parts \cite{2, 5}
\begin{equation}\label{eq:6}
\Delta S=\Delta_iS+\Delta_eS,
\end{equation}
in which $\Delta_eS$ is the entropy change due to the exchange of matter and energy with the environment and $\Delta_iS$ the entropy change due to "uncompensated transformation", the entropy produced by the irreversible processes in the interior of the system. $\Delta_eS$ equals $\dfrac{\langle Q\rangle}{T}$ where $\langle Q\rangle$ is the heat exchanged between the system and the reservoir \cite{2, 5}.
For any quantum dynamical process with $\dim(\mathcal{H})<+\infty$, the rate of the entropy change is given by \cite{6}
\begin{equation}\nonumber
\dfrac{dS}{dt}=-tr\{\dot{\rho_s}(t)\ln\rho_s(t)\}.
\end{equation}
Thus substituting $S(\rho)=-tr\{\rho_s(t)\ln\rho_s(t)\}$ into Eq. (\ref{eq:6}) then taking the time derivative of Eq. (\ref{eq:6}) we have
\begin{equation}\label{7}
-tr\{\dot{\rho_s}(t)\ln\rho_s(t)\}=\dfrac{d_iS}{dt}+\dfrac{\langle \dot{Q}\rangle}{T},
\end{equation}
where \cite{Alicki}
\begin{equation}\label{8}
\langle \dot{Q}\rangle\equiv tr\{\dot{\rho}_s(t)H_s\}.
\end{equation}
After some straightforward calculations we get
\begin{equation}\label{9}
\dfrac{d_iS}{dt}=tr\{(\dot{\rho_s}(t)\rho_s^\beta)(\dfrac{1}{\rho_s^\beta}(\ln\rho_s^\beta-\ln\rho_s(t)))\},
\end{equation}
where $\rho_s^\beta=\exp(-\beta H_s)/Z_s$ is the Gibbs state of the system. Now, analogous to De Donder's definition, we define the thermodynamic force and flow, respectively, as
\begin{equation}\label{10}
F_{th}\equiv\dfrac{1}{\rho_s^\beta}(\ln\rho_s^\beta-\ln\rho_s(t)),
\end{equation}
\begin{equation}\label{11}
V_{th}\equiv\dot{\rho_s}(t)\rho_s^\beta.
\end{equation}
\section{Supplementary Note 2}
Nielsen proved \cite{8} that a pure bipartite entangled quantum state $|\psi\rangle$ can be transformed into another pure bipartite entangled state $|\phi\rangle$ by local operations and classical communication (LOCC) if and only if $\alpha\prec\beta$, where the probability vectors $\alpha$ and $\beta$ denote the Schmidt coefficient vectors of $|\psi\rangle$ and $|\phi\rangle$, respectively. Here the symbol $\prec$ stands for the "majorization". We refer the reader to read Ref. \cite{8} and the references cited therein to read more about LOCC. We denote a quantum state by the probability vector of its Schmidt coefficients. An n-dimensional probability vector $\textit{x}$ is said to be majorized by another n-dimensional probability vector $\textit{y}$, written $\textit{x}\prec\textit{y}$, if the following relation holds:
\begin{equation}\label{14}
\sum_{i=1}^\ell x^\downarrow_i \leq \sum_{i=1}^\ell y^\downarrow_i   {\rm \ \ \ for\ any\ \ \ }1\leq\ell<n,
\end{equation}
where $x^\downarrow$ denotes the vector obtained by sorting the components of $x$ in nonincreasing order.
\begin{thm}\label{Theorem1}
(Theorem II.3.1 of Ref. \cite{9}). Let $x, y \in \mathbb{R}^n$. Then the following two conditions are equivalent:

$(\textit{i})\ \textit{x} \prec \textit{y}.$

$(\textit{ii})\ tr\varphi(x)\leq tr \varphi (y),$

for all convex functions $\varphi$ from $\mathbb{R}$ to $\mathbb{R}$, where $tr\varphi(x)\equiv\sum_{i=1}^n\varphi(x_i)$.
\end{thm}
Since $\bar{A}$ is a convex function, using this theorem and Nielsen's theorem \cite{8}, we conclude that the state $\rho$ can be transformed into the state $\sigma$ by LOCC if and only if $\bar{A}(\rho)\leq\bar{A}(\sigma)$.
\section{Supplementary Note 3}
Vidal \cite{12} discovered that there is always a maximal probability for incomparable states to be transformed into each other. Let $P(\ket{\psi} \rightarrow \ket{\phi})$ denote the maximal transformation probability of obtaining the state $\ket{\phi}$ from $\ket{\psi}$ by LOCC, then
\begin{equation}\label{eq:36}
P(\ket{\psi} \rightarrow \ket{\phi})=\min_{1\leq
\ell\leq n} \frac{E_\ell(\alpha)}{E_\ell(\beta)}=\dfrac{\alpha_n+\alpha_{n-1}+...+\alpha_{\ell}}{\beta_n+\beta_{n-1}+...+\beta_{\ell}}.
\end{equation}
where $n$ is the maximum of the Schmidt coefficients of $\ket{\psi}$ and
$\ket{\phi}$, and $E_\ell(\textit{x})$ denotes the abbreviation of
$\sum_{i=\ell}^n x^\downarrow_i$ for probability vector $\textit{x}$.
In the following we prove a theorem to show that the potential difference between two states predicts
which state is more probable to be transformed (or pulled) into another.
\begin{thm}\label{Theorem2}
Let $\ket{\psi}$ and $\ket{\phi}$ be two states with Schmidt numbers $\alpha$ and $\beta$, respectively. The $\ell$-th component of the potential difference $\triangle A_\ell=A_\ell(\rho_\psi)-A_\ell(\rho_\phi)>0$ if and only if the transformation $\ket{\phi} \rightarrow \ket{\psi}$ is more probable than $\ket{\psi} \rightarrow \ket{\phi}$ by LOCC.
\end{thm}
\textbf{Proof.} Consider the two states $\ket{\psi}$ and $\ket{\phi}$ with Schmidt numbers $\alpha$ and $\beta$, respectively. Thus the potential difference $\triangle A$ between these states reads,
\begin{equation}\label{15}
A(\rho_\psi)-A(\rho_\phi)=(\ln\dfrac{\beta_1}{\alpha_1},\ln\dfrac{\beta_2}{\alpha_2},...,\ln\dfrac{\beta_\ell}{\alpha_\ell},...,\ln\dfrac{\beta_n}{\alpha_n})\nonumber.
\end{equation}
Suppose there exist some $\alpha_j$, $\beta_j$ such that $\dfrac{\alpha_j}{\beta_j}<\dfrac{\alpha_\ell}{\beta_\ell}<1$ for $j\neq\ell$. Thus $\dfrac{\alpha_\ell}{\alpha_j}>\dfrac{\beta_\ell}{\beta_j}$. Rewriting Eq. (\ref{eq:36})
\begin{equation}\label{37}\nonumber
P(\ket{\psi} \rightarrow \ket{\phi})=\dfrac{\alpha_n+\alpha_{n-1}+...+\alpha_j+\alpha_{\ell}}{\beta_n+\beta_{n-1}+...+\beta_j+\beta_{\ell}},
\end{equation}
and since $\dfrac{\alpha_\ell}{\alpha_j}>\dfrac{\beta_\ell}{\beta_j}$, we must have
\begin{equation}\label{38}
\dfrac{\alpha_n+\alpha_{n-1}+...+\alpha_j}{\beta_n+\beta_{n-1}+...+\beta_j}<\dfrac{\alpha_n+\alpha_{n-1}+..+\alpha_j+\alpha_{\ell}}{\beta_n+\beta_{n-1}+...+\beta_j+\beta_{\ell}},
\end{equation}
which is a contradiction. Hence
\begin{equation}\label{39}
\dfrac{\alpha_\ell}{\beta_\ell}\leq\dfrac{\alpha_{\ell+1}}{\beta_{\ell+1}},... ,\ \dfrac{\alpha_{n-1}}{\beta_{n-1}},\ \dfrac{\alpha_n}{\beta_n}.
\end{equation}
Now according to Vidal's work \cite{12} if
$$P(\ket{\psi} \rightarrow \ket{\phi})<P(\ket{\phi} \rightarrow \ket{\psi}),$$
then $\ln\dfrac{\beta_\ell}{\alpha_\ell}$ is positive. This implies that if the transformation $\ket{\phi}\rightarrow\ket{\psi}$ is more probable than the transformation $\ket{\psi}\rightarrow\ket{\phi}$, we have
$A_\ell(\rho_\psi)>A_\ell(\rho_\phi)$.
The converse is also true. Hence, the $\ell$-th competent of the potential difference $\triangle A_\ell$, which is the largest difference between the components of the state potentials, determines in which direction the transformation is more probable. See the following example. Consider three states $\psi_k\in\mathcal{C}^4\otimes\mathcal{C}^4$, the square of the Schmidt coefficients of $k$-th state being $\vec{\alpha}_{k}$, where
\begin{eqnarray}\label{40}
\vec{\alpha}_{k=1} &\equiv& \frac{1}{122}(90, 12, 10, 10),\nonumber\\
\vec{\alpha}_{k=2} &\equiv& \frac{1}{122}(55,  55,  6, 6),\nonumber\\
\vec{\alpha}_{k=3} &\equiv& \frac{1}{122}(40, 40, 40, 2).
\end{eqnarray}
According to Vidal's theorem \cite{12}
\begin{eqnarray}\label{41}
P(\psi_1\rightarrow\psi_2)&=&\%32,\ P(\psi_2\rightarrow\psi_1)=\%60,\nonumber\\
P(\psi_1\rightarrow\psi_3)&=&\%39,\ P(\psi_3\rightarrow\psi_1)=\%20,\nonumber\\
P(\psi_2\rightarrow\psi_3)&=&\%28,\ P(\psi_3\rightarrow\psi_2)=\%33.
\end{eqnarray}
Now let us use quantum affinity to predict the results above. The potential differences $\Delta A$ for these transformations read
\begin{eqnarray}\label{42}
A(\rho_{\psi_2})-A(\rho_{\psi_1}) &=& (0.49, -1.52, 0.51, 0.51),\nonumber\\
A(\rho_{\psi_3})-A(\rho_{\psi_1}) &=& (0.81, -1.20, -1.38, 1.60),\nonumber\\
A(\rho_{\psi_3})-A(\rho_{\psi_2}) &=& (0.31, 0.31, -1.89, 0.69).
\end{eqnarray}
For transformations $\psi_1\rightleftharpoons\psi_2$, since $\Delta A_\ell=-1.52$ then the transformation $\psi_2\rightarrow\psi_1$ is more probable which completely agrees with the previous result. In the same way for transformations $\psi_1\rightleftharpoons\psi_3$ and $\psi_2\rightleftharpoons\psi_3$ we have $\Delta A_\ell=1.60$ and $\Delta A_\ell=-1.89$, respectively. Thus transformations $\psi_1\rightarrow\psi_3$ and $\psi_3\rightarrow\psi_2$ are more probable.
\section{Supplementary Note 4}
The entries of the thermal Gibbs state can be approximated with arbitrarily high accuracy with rational numbers as \cite{13}
\begin{equation}\label{16}\nonumber
\rho^\beta=(\dfrac{D_1}{D},..., \dfrac{D_d}{D}), \ D=\sum_{i=1}^{d}D_i,
\end{equation}
where $D_i, D \in \mathbb{N}$. A d-dimensional probability distribution $\textbf{p}$ is sent to a D-dimensional probability distribution $\hat{\textbf{p}}$ by an embedding map $\Gamma^\beta$ as follows \cite{13}
\begin{equation}\label{17}\nonumber
\hat{\textbf{p}}=\Gamma^\beta(\textbf{p})\equiv(\dfrac{p_1}{D_1},...,\dfrac{p_1}{D_1},..., \dfrac{p_d}{D_d},..., \dfrac{p_d}{D_d}).
\end{equation}
Now consider two states $\rho$ and $\sigma$ block diagonal in energy eigenbasis with probability vectors \textbf{p} and \textbf{q}, respectively. The necessary and sufficient conditions for block diagonal state interconversion under thermal operations is expressed as \cite{13}
\begin{equation}\label{18}
\mathcal E^\beta(\rho)=\sigma\ \ \textit{iff}\ \ \hat{\textbf{p}}\succ\hat{\textbf{q}},
\end{equation}
where
\begin{equation}\label{19}\nonumber
\mathcal E^\beta(\rho)=Tr_B[U(\rho\otimes\rho_B^\beta)U^\dag],
\end{equation}
with U satisfying $[U, H+H_B]=0$, $H$ Hamiltonian of the system and $H_B$ being arbitrary. Now using quantum affinity $A$ and Theorem \textbf{\ref{Theorem1}} we can state the following theorem as:
\begin{thm}\label{Theorem3}
A state $\rho$ block diagonal in energy eigenbasis can be transformed with certainty into another block diagonal state $\sigma$ by thermal operations if and only if
\begin{equation}\label{20}
\bar{A}(\hat{\alpha})>\bar{A}(\hat{\beta}),
\end{equation}
in which $\alpha, \beta$ are the probability vectors of the states $\rho$ and $\sigma$, respectively.
\end{thm}
\section{Supplementary Note 5}
Regardless of the physical details, the dynamics of open quantum systems can be roughly divided into two categories based on the memory effect of the reservoir \cite{3}: Markovian and non-Markovian dynamics. Here we examine the behavior of $A(\rho_s(t))$ during Markovian and non-Markovian evolutions and show how it acts as the thermodynamic force driving the flow and backflow of information. Any local-in-time master equation, for a quantum system having a $d$-dimensional Hilbert space, can be written in the form \cite{15}, in the interaction picture,
\begin{eqnarray}\label{23}
\dot\rho_s=\sum_{k=1}^{d^2-1}\gamma_k(t)[L_k(t)\rho_sL_k^\dag(t)-\dfrac{1}{2}\{L_k^\dag(t)L_k(t),\rho_s\}],
\end{eqnarray}
where the $L_k(t)$ form an orthogonal basis set of traceless operators, i.e.,
\begin{equation} \label{24}
tr[L_k(t)]=0,\ tr[L_j^\dag(t)\, L_k(t)] =\delta_{jk}.
\end{equation}
The dynamics is Markovian if and only if all decoherence rates $\gamma_k(t)$ are positive and correspondingly, non-Markovian when one or more of $\gamma_k(t)$ are negative \cite{15}. $A(\rho_s(t))$ is a function of the map $\Lambda_t$, thus it behaves differently during Markovian and non-Markovian evolutions. The collapses and revivals of $A(\rho_s(t))$ in the first example, below, show the fact that the dynamics of the system undergoes Markovian and non-Markovian evolutions during the process. But it should be noted that $A(\rho_s(t))$ is the total thermodynamic affinity at time t. Consider a master equation with two decay rates, $\gamma_1(t)>0$ and $\gamma_2(t)<0$, that is non-Markovian at all times. If the Markovianity dominates the non-Markovianity, i.e, $|\gamma_1(t)|>|\gamma_2(t)|$ then revivals are not observed in the behavior of $A(\rho_s(t))$, although the dynamics is non-Markovian at all times (see the second example). In order to separate the contributions of Markovianity and non-Markovianity in $A(\rho_s(t))$ we take the time derivative of $\bar{A}(\rho_s(t))$,
\begin{equation}\label{eq:25}
\dfrac{d\bar{A}(\rho_s(t))}{dt}=-tr\{\dot\rho_s(t)\rho^{-1}_s(t)\}.
\end{equation}
If we define
\begin{equation}\label{26}
\dfrac{d\bar{A}^k}{dt}\equiv-tr\{\gamma_k(t)[L_k(t)\rho_sL_k^\dag(t)-\dfrac{1}{2}\{L_k^\dag(t)L_k(t),\rho_s\}]\rho^{-1}_s\},
\end{equation}
Eq. (\ref{eq:25}) can now be written as,
\begin{equation}\label{27}
\dfrac{d\bar{A}(\rho_s(t))}{dt}=\sum_{k=1}^{d^2-1}\dfrac{d\bar{A}^k(\rho_s(t))}{dt}.
\end{equation}
Now the effect of Markovianity and non-Markovianity can be clearly seen separately.
\end{document}